MAJOR STEPS IN THE DISCOVERY OF ADIABATIC SHEAR BANDS

B. Dodd[1], S.M. Walley[2], R. Yang[3], V.F. Nesterenko[4]
[1] The Institute of Shock Physics, Imperial College, London, SW7 2BW UK
[2] SMF Fracture and Shock Physics Group, Cavendish Laboratory, Cambridge CB3 0HE, UK
[3] Laboratory of Nonlinear Mechanics, The Institute of Mechanics, Chinese Academy of Sciences, Beijing 100190, P.R. China
[4] Jacobs School of Engineering, Department of Mechanical and Aerospace Engineering, University of California at San Diego, La Jolla, CA 92093-0403, USA

When one of us (BD) co-authored with Prof. Yilong Bai a book on adiabatic shear bands (ASBs) in 1992 [1], the standard story of the discovery of ASBs was believed to have started with the American researchers Zener & Hollomon's famous 1944 paper in the *Journal of Applied Physics* [2] where the phenomenon was first reported and named. The story then moved back to France in the 1870s to Henri Tresca's reports on heat crosses seen in the forging of platinum alloys [3, 4]. Tresca's study was repeated in England by Massey in 1921 using steel [5] but the first photograph of the heat cross phenomenon was taken by Johnson and co-workers in 1964 [6], again using steel.

A close reading of Zener & Hollomon's 1944 paper reveals that although they used the three words 'adiabatic', 'shear', and 'band' in their paper, they did not put the three words together to form a single phrase. However, there can be no doubt that they observed ASBs, as can be seen from figure 1, reproduced from their paper. It is a moot point whether the heat crosses reported earlier by Tresca and Massey should be called 'adiabatic shear bands' as the shear localization they saw is much more diffuse than that shown in figure 1, as can be seen from the photograph of the phenomenon taken later by Johnson and co-workers (figure 2).

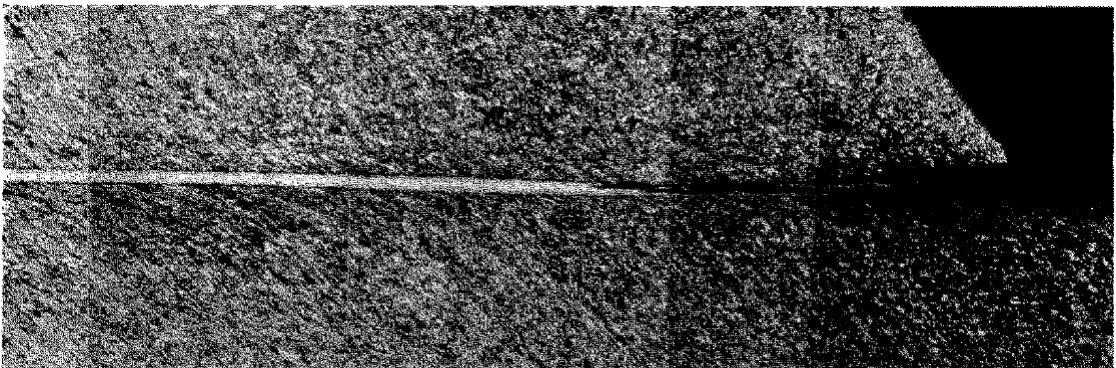

*Figure 1. ASB formed by the dynamic punching of a steel plate using a drop-weight machine. From [2].*

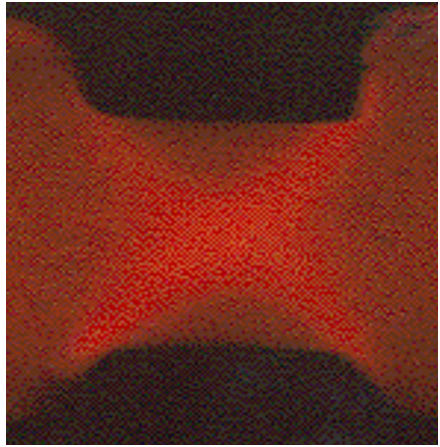

*Figure 2. Photograph of the heat cross in mild steel produced by the punch forging of mild steel. This study was performed by Johnson and co-workers in the same forging shop as Massey used in 1921. From [6].*

And so the story rested, and indeed was repeated by one of us in a review paper published in 2007 [7]. However, in August 2013, one of us (SMW) found a paper in the Cambridge University Library by Davidenkov & Mirolubov (of the Leningrad Institute of Physics and Technology) entitled: 'A special kind of upset deformation of steel: The Kravz-Tarnavskii effect' [8]. This paper is dated 1935 and was published in German in a major Soviet Journal aimed at non-Russian speakers (*Technical Physics of the USSR*). The editor of this journal was A. Joffe, the father of Soviet physics, and the editorial board included many other major Russian physicists of that time.

Tantalizingly, the 1935 paper is sub-titled 'The Kravz-Tarnavskii Effect' and as you will see from the figures reproduced from this paper (figures 3-5), the 'K-T Effect' is in fact adiabatic shear banding. The authors of the 1935 paper refer to a date of 1928 for the first publication of the phenomenon by Kravz-Tarnavskii in *Zeitschrift der Russischen Metallurgischen Gesellschaft* [9].

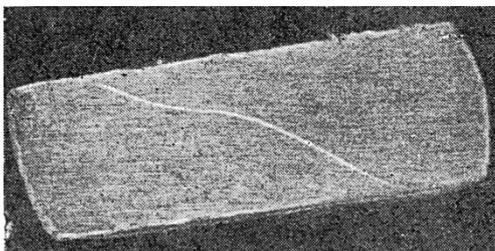

*Figure 3. Shape of band on an etched section of a steel specimen impacted by a 50 kg weight dropped from a height of 2.55m. From [8].*

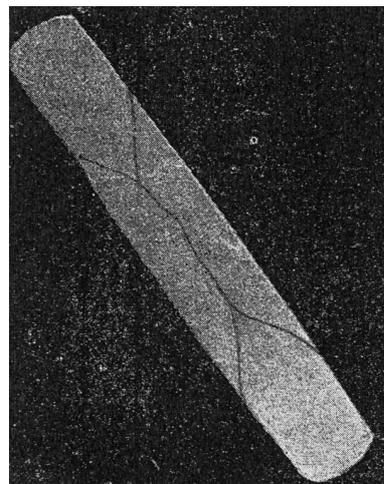

*Fig. 4. Form of the band with some branches. From [8].*

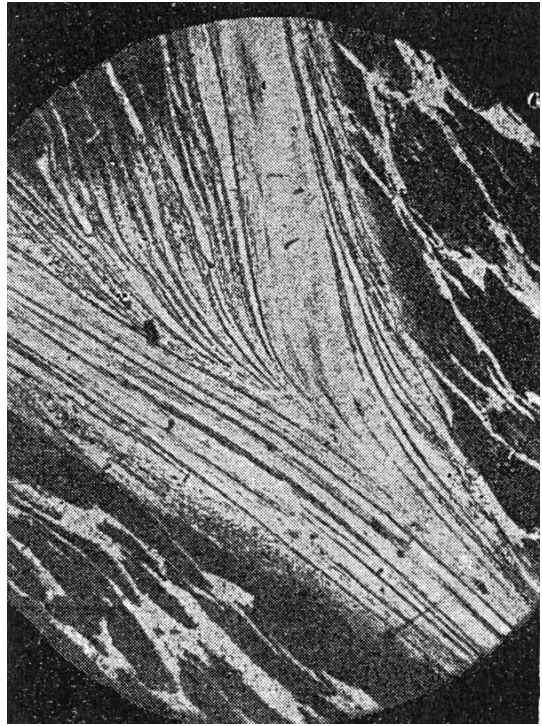

*Fig. 5. A band with ferrite penetrating into it. Steel with 0.66% C; initial structure: pearlite in ferrite; weight 50kg, height 4m. From [8].*

We were then stuck because there was no record of a journal bearing that name in any library in the United Kingdom or the United States. With hindsight, our assumption that the journal had a German title was wrong (it has a Russian name and was published in Russian). What about Russia itself? How could we perform a search of Russian libraries? Had all copies of the journal printed in 1928 been destroyed during the 20$^{th}$ century?

Our initial attempts to find the journal containing Kravz-Tarnavskii's paper in Russian libraries by asking for help from Russian experts in this research area were unsuccessful. Moreover, even in the Russian scientific literature, we found only a few examples where references to Kravz-Tarnavskii's and Davidenkov & Mirolubov's papers were made. This is very unusual especially because the phenomenon of shear localization first reported in Kravz-Tarnavskii's 1928 paper was later named after him in 1935 in a major Soviet scientific journal. Moreover the giving of his name to the phenomenon (which is not very often done in the scientific literature and in general only very reluctantly) occurred in a paper co-authored by N.N. Davidenkov who later became a Fellow of the Ukrainian Academy of Sciences. He published many papers in the international literature on metallurgy and he also wrote a book on the dynamic mechanical behaviour of metals [10]. But the sad fact is that the two papers by Kravz-Tarnavskii and Davidenkov & Mirolubov, reporting a new phenomenon and its extensive study, were practically forgotten in the scientific literature. This was probably due to the fact that Kravz-Tarnavskii's 1928 paper was published in a Russian journal which would have had limited circulation due to the lack of paper in Russia at that time[*]. Also Davidenkov & Mirolubov's 1935 paper was published in German (a relatively common practice in the 1930s) even though most papers in *Technical Physics of the USSR* were published in English.

However we were lucky because a friend of one of us (VFN) found the journal in the central Moscow library in March 2014 and was allowed to photograph Kravz-Tarnavskii's 1928 paper! Rong Yang then translated the Russian text into English, with VFN then checking and correcting the translation. The paper is entitled 'A peculiar band discovered in steel'. Rong Yang also performed a literal translation of Davidenkov and Mirolubov's 1935 paper written in German. The translation was then revised by BD and SMW into a more readable form of English and then checked for accuracy by one of our German colleagues, who remarked that the sentence structure was very difficult to understand even for a native speaker of German. We hope in due course to fully republish these two papers in English translation to make the content of these 'lost and found' papers available to all researchers working in the field of ASBs.

A selection of photographs from the 1928 Kravz-Tarnavskii paper are given in figures 6-8. These are believed to be the first published photographs of the ASB phenomenon. The figure captions in the original paper simply gave the magnification of the original photographs, and this has been respected in their reproduction in this article.

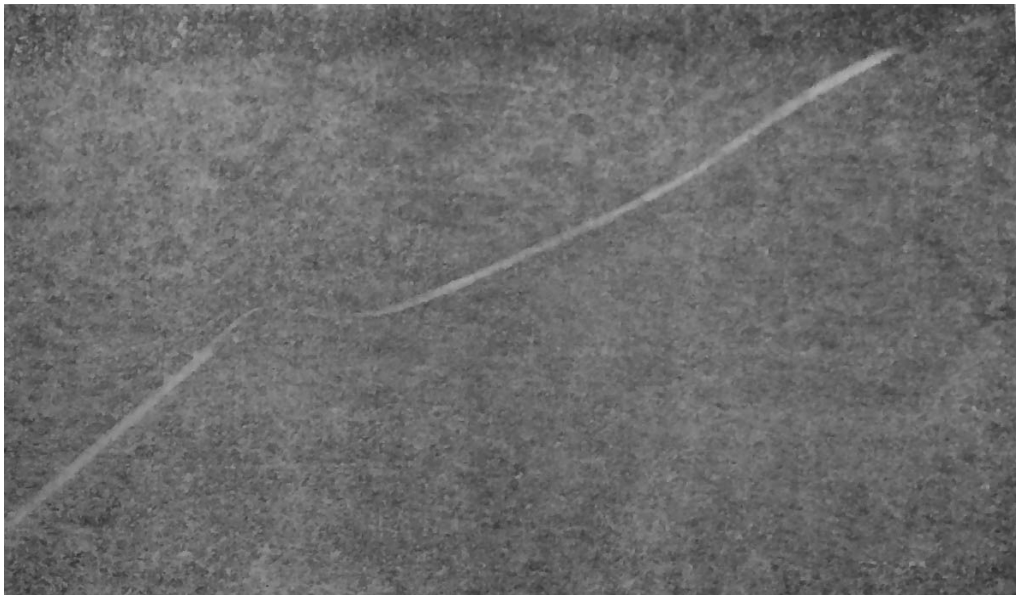

*Figure 6. Magnification 15X. From [9].*

---

[*] For those who are interested in the social background to performing research in Petrograd (then named Leningrad, now St Petersburg) at the time of the publication of Kravz-Tarnavskii's paper, V.F. Nesterenko recommends Ayn Rand's semi-autobiographical novel *We the Living*, published in 1936 in the USA.

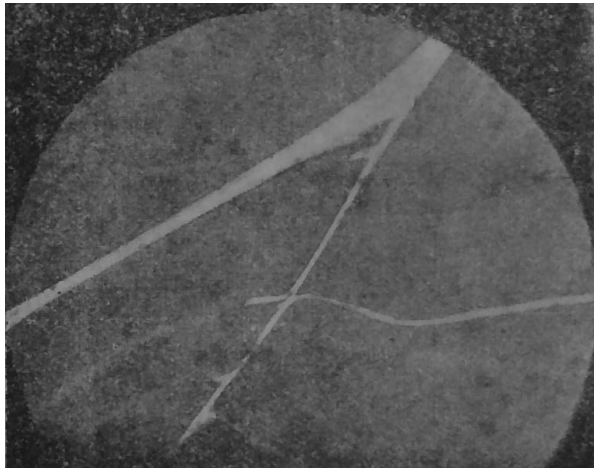 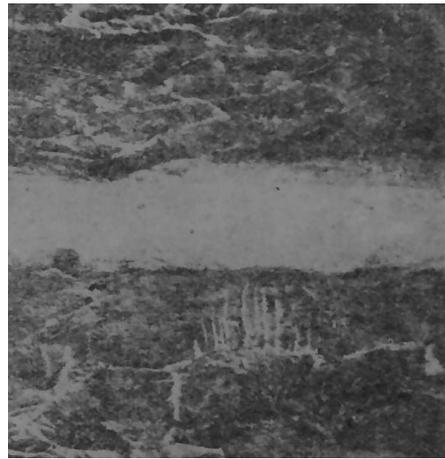

*Figure 7. Magnification 50X. From [9].*  *Figure 8. Magnification 150X. From [9].*

These two papers make it clear that the history of the discovery of ASBs should be revised as follows:

1. 1870s: Tresca mentioned that during forging plastic flow can localize into the shape of an X and that during the dynamic plastic deformation heat was generated [3, 4]. Massey in England subsequently published a study of these heat crosses in 1921 [5] but it is clear from the repeat of Massey's experiments carried out by Johnson and co-workers in the 1960s [6] that the localization of strain is more diffuse than in true ASBs. The X-shaped region of high temperature can be interpreted as the manifestion of the largest shear strain. This is expected to lie at around 45° to the impact direction and under some conditions may be the precursor to the formation of true ASBs.

2, 1928: Kravz-Tarnavskii first described true ASBs generated in specimens of various types of steel by drop-weight impact at 10 m/s [9]. He made the following important statements based on his observations:
(a) quasistatic deformation never resulted in shear localization;
(b) dynamic deformation produced by impact was a necessary condition for observation of this 'peculiar band' (shear localization zone) in different steel types with carbon content less than 0.8% and other special (though unspecified) steels;
(c) occurrence of this 'peculiar band' resulted in low energy absorption by the rest of the specimen, the energy of deformation being mostly absorbed by the 'peculiar band';
(d) the 'peculiar band' had a microstructure very different to the initial microstructure of the steel, in some cases being practically non-etchable, and with a high scratch hardness;
(e) he suggested that mass transfer (expulsion of ferrite from the shear localized zone) and subsequent refinement of the microstructure by fragmentation of brittle components was responsible for its unusual properties.
We would like to comment at this point that it is unclear how to accomplish mass transfer from a localized shear band which would involve normally slow diffusion processes unless the diffusion is greatly accelerated in the strongly sheared material and also in material nearby. But it should be noted that very finely fragmented microstructure was observed some years later inside localized shear bands in brittle granular materials [11].

3. 1934: Taylor & Quinney found experimentally that the amount of heat produced by plastic deformation was about 90% for most metals [12].

4. 1935: Independently Davidenkov and Mirolubov performed extensive experiments on a variety of steels with various microstructures and under different conditions of impact (energy and velocity) in collaboration with Kravz-Tarnavskii (even using some of his samples) [8]. They obtained many ASBs under certain impact conditions. They measured the thermal energy dissipated in the deformed samples with and without shear bands and observed no difference between corresponding values. They proposed two different mechanisms of shear localization according to their observation of the dynamic behaviour of steels processed under various conditions to produce different initial microstructures in order to explosive the role of microstructure on the formation of localized shear:
(a) Plastic deformation first occurs in the plane of maximum shear stress accompanied by local heating. Heating lowers the yield stress, so hardening is followed by weakening and further localized deformation continues along the same plane rather than in other regions;
(b) Microfracture along the plane of maximum shear stress forms an inclined crack, breaking the specimen into two parts. Subsequent sliding in the plane of the crack results in localized heating so that the crack welds itself together. This is followed by quenching of the heated metal layer which undergoes an austenitic transformation as it is quenched. Based on temperature estimates, they suggested that localized heating followed by rapid quenching due to the surrounding cold bulk metal may result in a martensitic transformation inside the shear band. The quenched microstructure in the localized shear band is different to the normal martensite formed by quenching. They suggested that the reason why the characteristic needle-like martensite micrsotructure was not formed was due to the high pressure and high rate of transformation.

They did not confirm the process of expulsion of ferrite from localized shear zones suggested in Kravz-Tarnavskii's earlier paper [9]. The studies by Davidenkov & Mirolubov are quite sophisticated, and even show side-pressings as reported much later in the 1980s by an American team. So at last we now know of the 'lost work' carried out in Russia on the ASB phenomenon.

5. 1944: Zener and Hollomon published a famous paper in the *Journal of Applied Physics* [2] containing photographs of ASBs generated under different loading conditions to the Russian studies. Zener & Hollomon's method consisted of dropping a weight upon a punch in contact with a plate (this may be called 'forced' shear localization). They suggested a mechanism of shear localization similar to the one first proposed by Davidenkov & Mirolubov emphasizing that: "a negative stress-strain curve implies an intrinsic instability of the material where deformation cannot be homogeneous; for a region which, by chance, suffers more deformation than the surrounding region and becomes weakened thereby, will continue to deform while the surrounding region undergoes no further strain". They also suggested that the white band of metal joining the punch and plate is martensite resulting from rapid quenching of material inside highly sheared material (shear strain of nearly 100).

We hope that the major steps in the discovery of the ASB phenomenon presented in this article places the efforts of researchers in different countries in a new historical perspective. Now proper tribute can be paid to the Russian pioneers.

**Note on translation policy**: We have used the English convention for transliterating Russian names, rather than the German one. Thus 'Kravz-Tarnavskii' rather than 'Krawz-Tarnawskij' (which is how his name appears in the sub-title of the 1935 paper, written in German).

**Appeal**: if anyone knows of a library containing the *Journal of the Russian Metallurgical Society*, it would be good to arrange for it to be scanned and indexed as there may not be many copies left in the world (it seems there is only a single copy left in Russian libraries). Who knows what gems of metallurgical research are waiting to be rediscovered in its pages.

**Acknowledgement**
The Institute of Shock Physics acknowledges the support of the Atomic Weapons Establishment, Aldermaston, UK and Imperial College, London.

A peculiar band discovered in steel
V.P. Kravz-Tarnavskii
Originally published in Russian in:
*J. Russ. Metall. Soc.* (1928) (3) 162-167

The question I wanted to find the answer to is: what are the microstructural changes in steel that can be produced by impact? These experiments were carried out by the impact of a drop hammer with a mass of 25 kg from a height of 5m onto a piece of metal with a volume of about 1 cm$^3$. Investigation of the samples which had been heavily deformed between the hard flat striker and the anvil revealed a phenomenon that had not been observed before.

In this preliminary report I present only a qualitative description of the phenomena produced by a single impact on specimens of medium carbon steel (C 0.39-0.49%). In all cases, the original structure of the samples consisted of sorbite in a large ferrite network. All the microstructures presented in this report were revealed by etching with a 4% solution of nitric acid in ethyl alcohol.

Under the specified impact conditions, there was a volume inside the specimens of about 0.01 cm$^3$, characterized by peculiar properties.

Cutting the samples in different directions revealed that this volume consisted of a band perpendicular to the direction of impact. The band in the deformed sample did not quite reach the surfaces parallel to the direction of impact; there were almost always one or two of these bands. The thicknesses of the bands were about 0.1 mm. Sometimes, the band split into a few other bands. Those that reached the outer surface of the specimen produced discolouration; they could be seen with the naked eye on sectioned specimens, even without polishing.

Sections of the specimens etched with conventional reagents and then polished clearly reveals the shiny band as it not affected even by prolonged etching.

The polished and etched sections (originally microstructure is a sorbite) are shown in Figs. 1 and 2. The appearance of the band is accompanied by significantly less heating of the sample under impact than when no band appeared. In the former case, the impact energy was absorbed mainly by the band.

As shown in Fig. 1, the entire specimen volume is not equally deformed; there are areas which have been subjected to negligible plastic deformation, which can be judged based on the microstructure of the ferritic network before and after impact. The band has heavily deformed zones on either side of it, but there are exceptions, see Fig. 3 which shows a ferrite scallop.

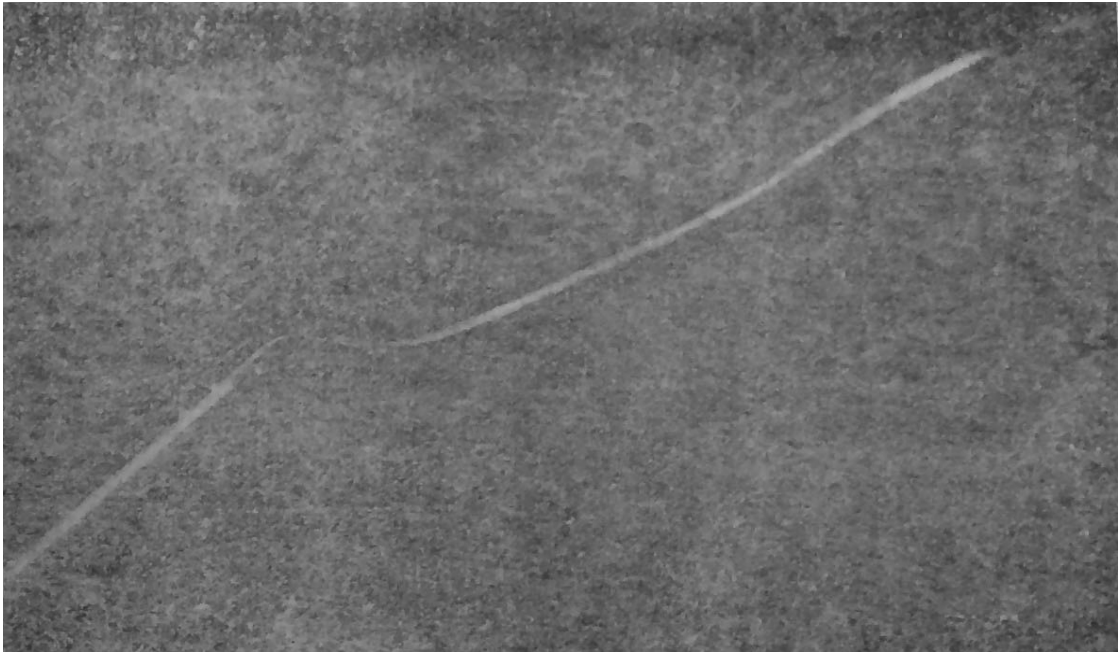

*Figure 1. Magn. 15X*

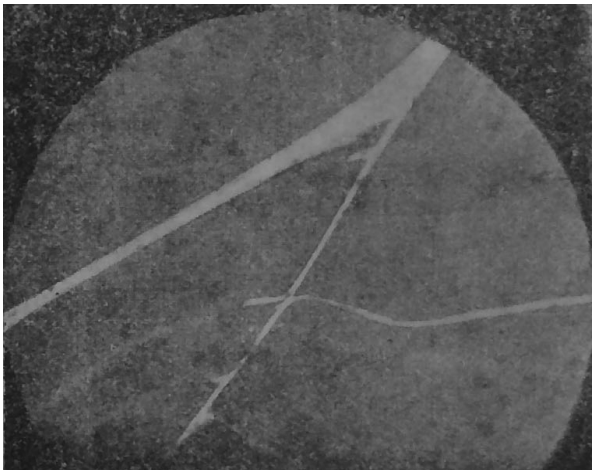

*Figure 2. Magn. 50X*

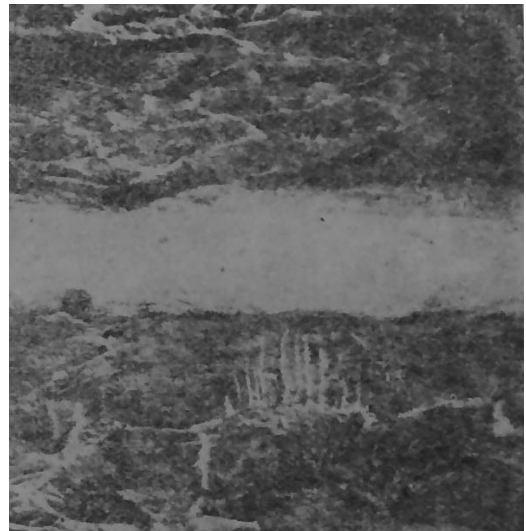

*Figure 3. Magn. 150X*

The bands did not only appear in medium carbon steel with this microstructure. I have obtained the same band structure in other carbon steels with up to 0.8% carbon, and also in a number of special steels with different microstructures.

Numerous attempts were made to obtain this peculiar band phenomenon under quasistatic compressive loading of the same steel grades, but no bands were ever observed.

The bands were very hard and their hardness was very different to the surrounding material. Measurement of the Martens [scratch] hardness using a load of 20g gave a scratch width inside the band of 0.0068 mm and 0.011 mm outside the band.

Under the microscope, the band appears as a non-etching white band, whilst the other structures (sorbite, ferrite or pearlite) remained the same as they were before the impact. The bands sometimes intersected an outer surface as shown in Fig. 4. Even at a magnification of 3500 the microstructure of the band was unresolvable (figure 5), even after strong etching.

Sometimes part of the ferrite network encroached inside the bands. This ferrite entry into the band often ended with dark 'dashes' (figure 6), having the appearance of cavities that were not illuminated by the lighting system used. Also these lines characteristically ended abruptly.

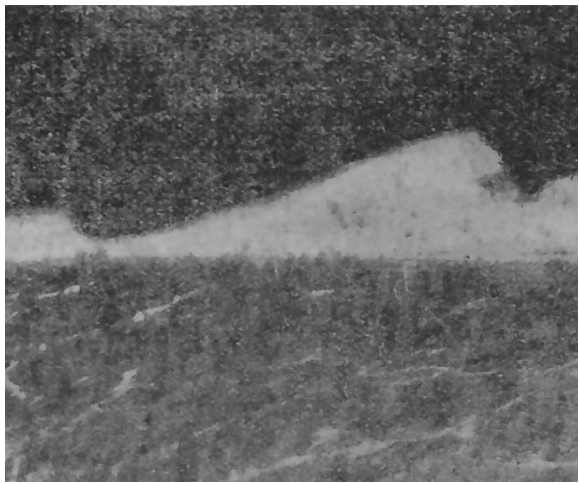

*Figure 4. Magn. 150X*

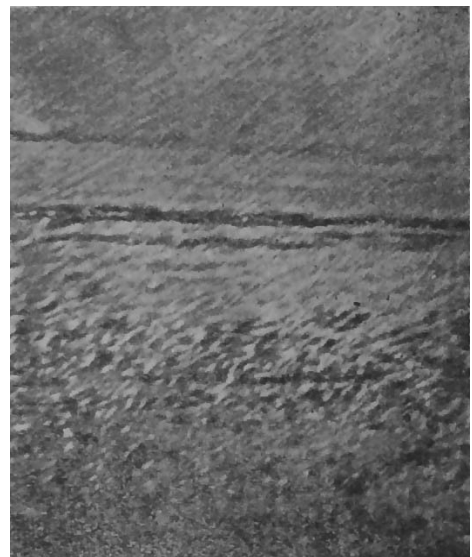

*Figure 5. Magn. 3500X*

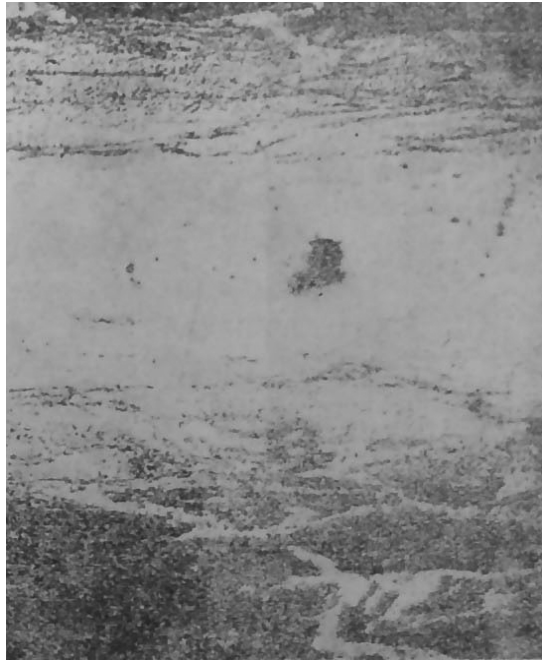
*Figure 6. Magn. 270X*

Annealing to a temperature of between 250 and 275 °C had no influence on band etching. Annealing at a higher temperature made the bands more etchable. When the temperature was between 650 and 680 °C a structure similar to fine sorbite was obtained. Fig. 7 shows the microstructure of the band after annealing at 650 °C. Fig. 8 shows the microstructure after annealing at 680 °C for the case when the original microstructure of the sample consisted of pearlite in a ferrite matrix.

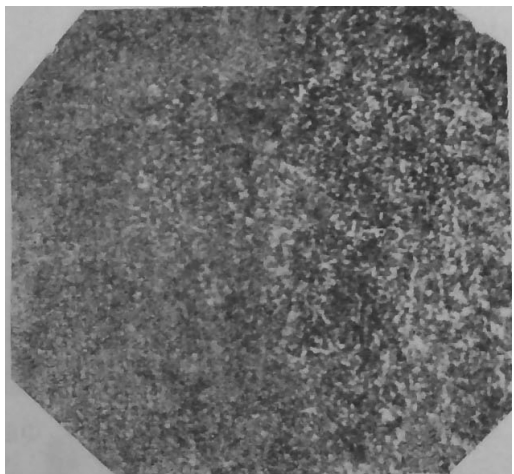  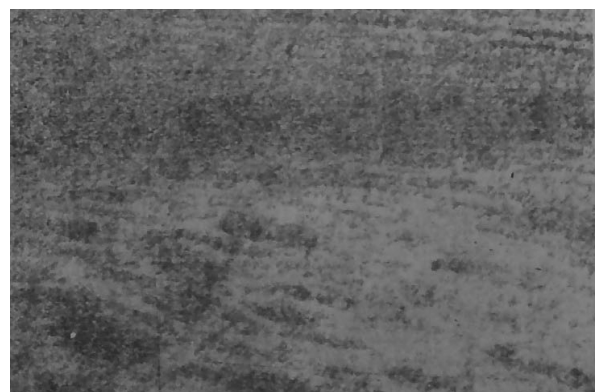
*Figure 7. Magn. 1100X*               *Figure 8. Magn. 270X*

Figure 9 presents the band microstructure after annealing at 700 °C for 20 minutes with subsequent cooling in air for the case where the original microstructure of the sample also consisted of pearlite in a ferrite matrix.

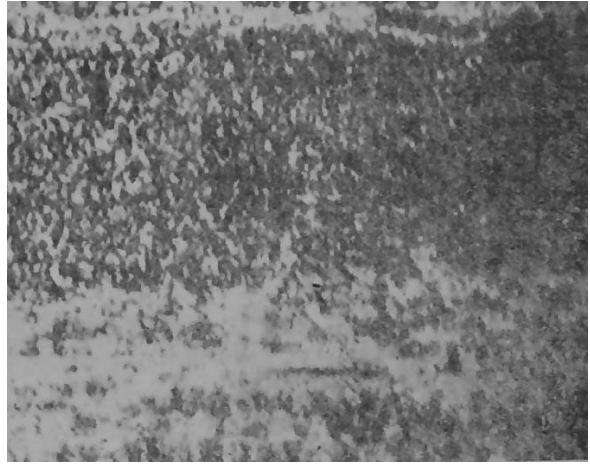

*Figure 9. Magn. 1600X*

Heating the band to 730 °C for 5 minutes in a lead bath and then air-cooling did not dissolve it. In this case the band microstructure, apparently, is identical to the initial sorbite but the bands had a slightly darker colour. The ferrite from the network which existed in the structure in the bulk before heat treatment, was preserved after the treatment, while this ferrite did not appear in the band. This pattern is shown in Fig. 10 where the upper part is a band, beneath it is the ferrite from the original network, and behind this is the original sorbite grain.

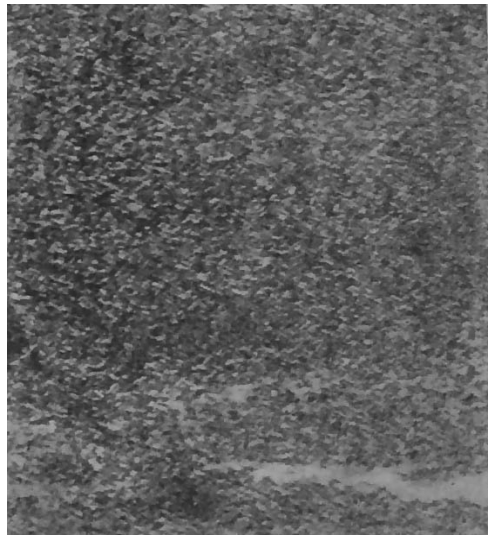

*Figure 10. Magn. 780X*

Annealing at a temperature 30-40 °C above $A_{c1}$, holding at this temperature followed by slow cooling, produced absorption of the bands so that afterwards only fragments of the etchable bands could be observed.

Depending on the chemical composition, the microstructural state of the sample and the impact conditions, the band that formed might be completely or partially etchable as shown in Figure 11, where the initial microstructure was pearlite lamellae in a ferrite matrix.

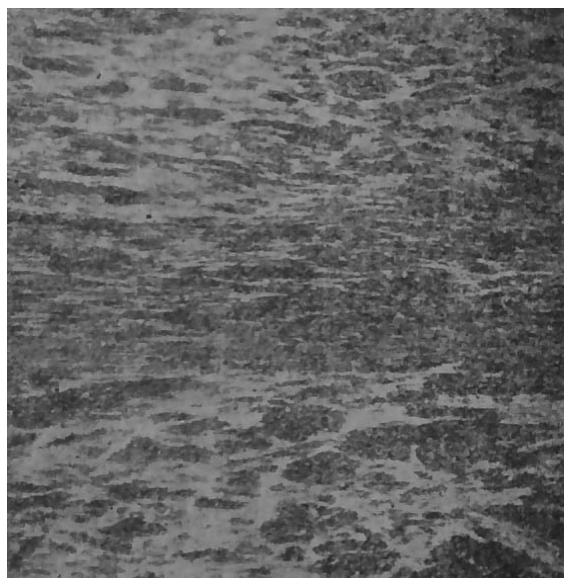

*Figure 11. Magn. 150X*

The ability to follow the bands in various stages of their formation, not only on the surface but also inside specimens, allows us to identify the stages of their formation and facilitates a study of their properties.

Analysis of the photographs presented in this paper of various cut sections suggested the hypothesis that the soft ferrite is 'squeezed' out of the band. The harder constituents richer in carbon therefore can move closer to each other and as a result the volume that absorbs the majority of the impact energy becomes 'cemented'. Due to this mechanism the band is enriched with carbon in comparison with the bulk of the sample. This carbon enrichment might continue up to the carbon content in the hard constituent or even higher[*].

---

[*] Note added in 2014 by V.F. Nesterenko: This is a bold hypothesis and probably the first attempt to involve the mechanical process of plastic deformation in redistribution of components during high strain rate shear flow through effecting phase transformation or chemical reaction (the relevant current area of research is mechanochemistry). But eventually their mechanism requires some diffusion process to redistribute components, which is normally a rather slow process. The subsequent study by Davidenkov and Mirolubov (1935) did not confirm this carbon enrichment mechanism (see the accompanying paper in translation).

The next step in band formation is the transformation of this 'cemented' volume into a very hard but previously unknown state, with probably a very finely fragmented microstructure[†].

I will continue further study of these bands and the peculiar properties of the metal inside them, which appears to be extremely hard.

I wish to thank I.I. Bokova who assisted me in this work and took all the photomicrographs.

---

[†] Note added in 2014 by V.F. Nesterenko: Fragmentation of brittle constituents, as an alternative softening mechanism to thermal softening inside shear bands, was a reasonable hypothesis and has subsequently been seen in shear bands in granular materials.

A special kind of upset deformation of steel: The Kravz-Tarnavskii Effect
N. Davidenkov & I Mirolubov
Originally published in German in:
*Tech. Phys. USSR* **2** (1935) 281-298

In 1928, Kravz-Tarnavskii published a report in the *Journal of the Russian Metallurgical Society* with the title "A peculiar band discovered in steel" [see companion paper]. In this article, he reported that a peculiar microstructural change occurred in steel when it is subjected to [dynamic] compression. This change had not previously been observed by anyone and therefore rightly deserves to be called 'the Kravz-Tarnavskii Effect'. The purpose of our study was to investigate the conditions under which the Kravz-Tarnavskii phase forms and to determine its mechanical properties. The experimental programme was designed in collaboration with Kravz-Tarnavskii, who also made the specimens for our study. However, our interpretations of the experimental results differ from those of Kravz-Tarnavskii.

**Description of the phenomenon**.
In his 1928 paper, Kravz-Tarnavskii reported that he had performed drop-weight compression experiments on steel specimens approximately 1 cm$^3$ in size. These experiments produced peculiar bands about 1 mm in thickness, inclined at an angle to the direction of impact (Fig. 1). The bands sometimes propagated to the curved surfaces of the specimens, where they exhibited different geometric dimensions (Fig. 2). The direction of the bands almost coincided with the diagonal of the original specimen (which had a rectangular parallelepiped shape). However, the bands were not straight: they had one, two or more bends in them (Figs. 1 & 3). The place where the bands intersect the specimen surface is almost always tarnished, the colour varying from straw-yellow to blue[*] (Figs. 2, 4). Kravz-Tarnavskii found that the appearance of the band depended on the impact of the specimen, which was either not heated or heated considerably less than when no bands were formed. Polishing and etching a cross-section with the usual reagents revealed a shiny, non-etchable band, surrounded by the initial microstructure of the steel (Fig. 5). However, for certain chemical compositions and microstructures, bands formed which may be etched slightly (Fig. 7) i.e. they were not fully formed.

---

[*] Note from the translators: these are the well-known steel tempering colours. For a recent example obtained in a similar experiment, see Walley *et al.* (2006) *J. Phys. IV France* **134** 851-856

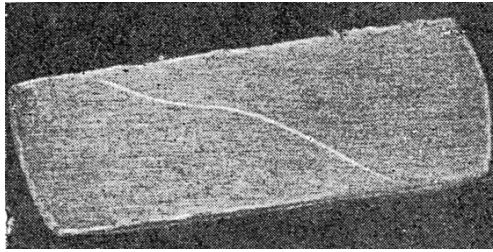

*Fig. 1. Shape of band on an etched section of a specimen impacted by a 50 kg weight dropped from a height of 2.55m.*

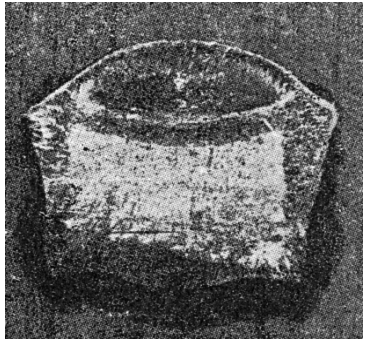

*Fig. 2. The surface of an impacted specimen. The region towards the top of the photograph is where the band exited the specimen and also where the specimen stuck to the drop-weight. Below the spot is a dark stripe due to visible tarnishing.*

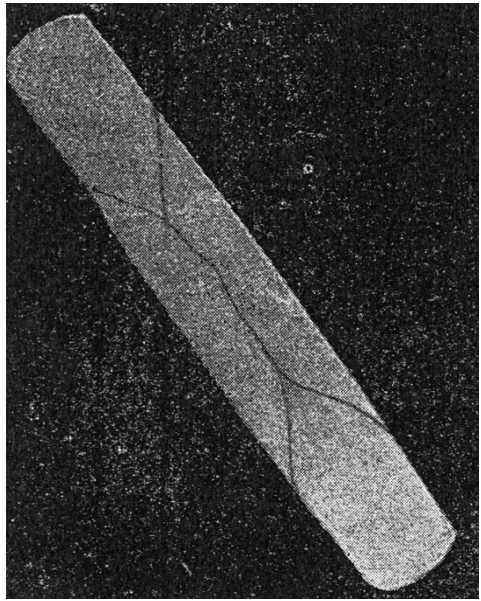

*Fig. 3. Form of the band with some branches.*

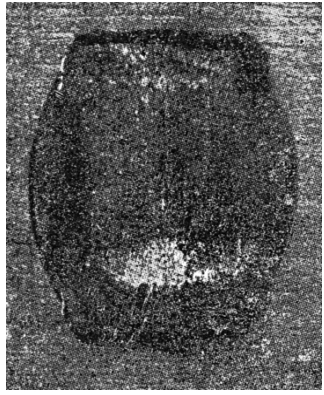

*Fig. 4. The surface of an impacted specimen. The dark regions at top and bottom are untarnished.*

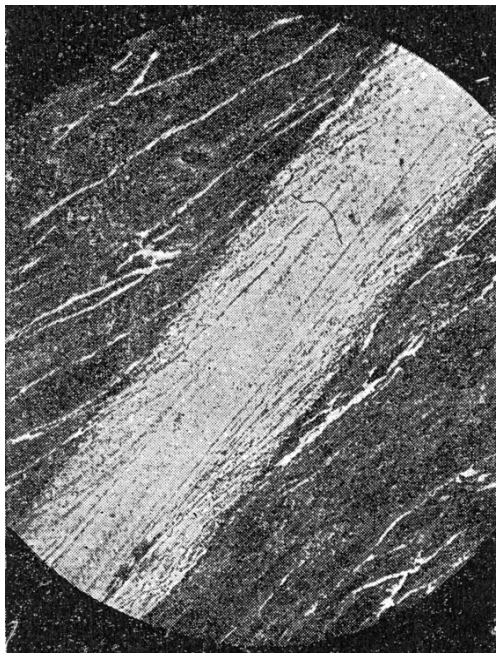

*Fig. 5. Microstructure of a band. Steel with 0.39% C; first state-sorbite in ferrite; impact produced by a 50 kg dropweight falling from a height of 2.55m.*

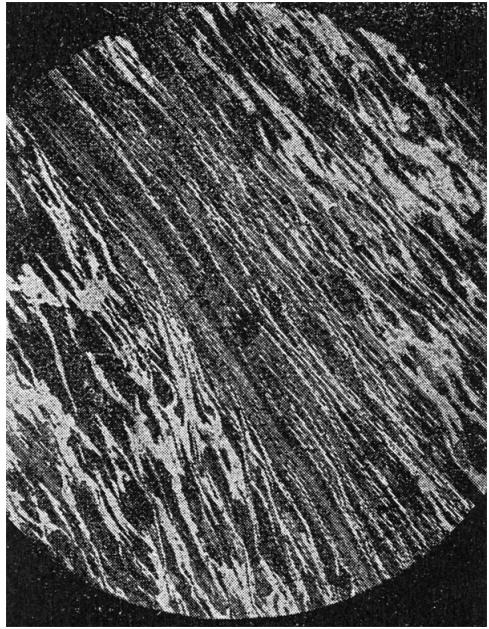

*Fig. 6. A band. Steel with 0.39% C; initial structure: pearlite in ferrite; weight 50kg, height 2m.*

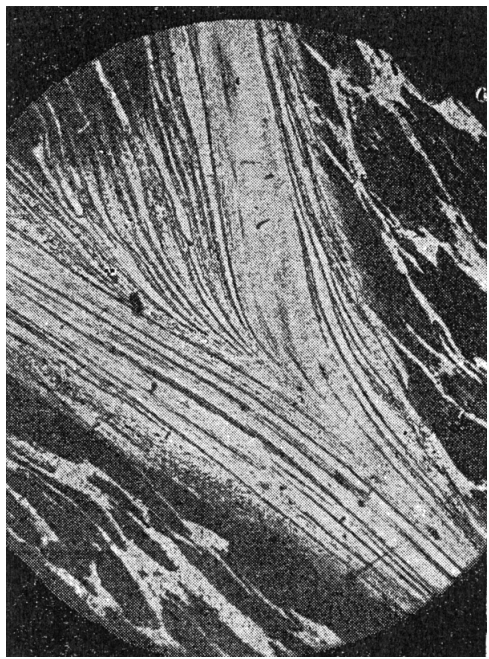

*Fig. 7. A band with ferrite penetrating into it. Steel with 0.66% C; initial structure: pearlite in ferrite; weight 50kg, height 4m.*

The microstructure of the metal outside the band remained the same as before impact. The lamellar pearlite and sorbite grains in the ferrite were unchanged (Fig. 7). Sometimes ferrite partially penetrates the band, or remains in the band zones, where the ferrite was previously (Fig. 5, 7).

The hardness of the band was extremely high, differing sharply from the hardness of the surrounding material. It is impossible to produce the band by quasistatic compressive loading. Annealing at temperatures of between 250˚-270˚C had no influence on the metallographic properties of the band. When subjected to a further annealing at higher temperatures, a structure formed which is very similar to fine sorbite. Annealing above the lower critical point for a longer time followed by slow cooling caused the bands to disappear: only their outline remained visible.

As discussed before, the assumption was made (based on the previous study by Kravz-Tarnavskii) that the solid components of the relatively soft ferrite gradually approached each other and were removed from the volume occupied by the band. As a result, the band was enriched in carbon. This enrichment may be up to the carbon content of the harder component, or to an even higher level. In this way, the cemented volume became a very hard, extremely finely divided phase, which determined the nature of the band.

**Problem definition.**
Although by this point we have set out a significant number of facts, we cannot regard them as adequate for choosing one theory over against another for understanding the physical nature of the band. Therefore, we decided, in agreement with Kravz-Tarnavskii, to perform some additional experiments, which could shed light on the recently observed phenomenon.

To set the trajectory for the experiments, we made the following assumptions about the formation of the band, which seemed to us the most likely working hypothesis.

As a consequence of the localized deformation, damage and crushing occurred along a slip plane in the material. Here very fast slip developed at first, releasing a large amount of heat (due to the strong friction), which then diffused into the main body of the specimen, the steel anvil and the drop-weight. Therefore at the place of deformation localization, the temperature probably exceeds the critical point for transformation to austenite. It then undergoes intense quenching. The material of the band itself is in the martensitic state. The only reason why no characteristically needle-like microstructure formed was because it underwent a special, not yet investigated condition of high pressure and high rate of conversion.

It is very likely that for this process to occur a high temperature is required in the slip plane. This was supported by the following facts: (i) colours appeared on the surface due to tarnishing, and (ii) the sample stuck to the anvil at these sites (Fig. 2).

A simple calculation on the basis of the observed temperatures shows that it is sufficient for a volume one sixth the size of the specimen to be heated to 1500 ˚C during impact. But the volume of the band is only 1% that the specimen. Therefore, the material in the band must reach a temperature far exceeding the critical point.

As for heat-dissipation, the quenching rate must be very high, since the high-temperature band was contiguous with the large mass of surrounding cold steel, a high thermal conductivity material.

**Experiments and results**
*1. The heating of specimens produced by impact*
In his original observations Kravz-Tarnavskii noticed that the specimens heated differently depending on whether a band formed in them or not. Since he evaluated the heating of the specimens simply by touching them, we decided it would be interesting to determine the differences in the temperatures by exact measurement.

For this purpose, 10 specimens of steel with a carbon content of 0.39% and dimensions 13x13x7 mm$^3$ were subjected to impact by a 50 kg drop-weight dropped from a height of 2.55m using the vertical Amsler striking mechanism. The initial structure of the specimens was sorbite in ferrite. The heat dissipated in the specimen was determined using a specially made small brass calorimeter into which a copper-constantan thermocouple was inserted, connected to a mirror galvanometer.

After the impact and rebound of the drop-weight, the specimen was pushed as quickly as possible from the anvil into the calorimeter using a wooden stick. The water was then stirred and the maximum deflection of the galvanometer observed. For each new test, the water was replaced (its volume was held constant at 15 cm$^3$) and the galvanometer was reset to zero. In order for the specimen not to bounce off the anvil after impact, a paper plate was attached to the latter.
The results of these tests are given in Table 1.

It can be seen from the table, in contrast to the subjective results previously obtained by simply touching the specimens, that to within experimental error the same amount of heat was dissipated in the specimens in which bands formed as in those where no bands were seen.

A simple calculation shows that the temperature which the whole steel specimen should be theoretically heated to for the given impact energy (50 x 2.55 kg) is about 300 °C, requiring a heat of 292 kcals. The experiment showed, however, that the specimen had a temperature after impact of about 120 °C, corresponding to a quantity of heat of about 107 kcals. It can be seen, therefore, that about 63% of the total energy dissipated in the specimen was conducted into the anvil and the drop-weight or stored as potential energy of internal stress.

**TABLE 1**
*Heat measurements in band formation*

| Specimen no. | Drop height / cm[*] | Galvanometer deflection after impact | Band formed? | Middle galvanometer deflection | Mean absolute error | Mean percentage relative errors | Percentage experimental error |
|---|---|---|---|---|---|---|---|
| 1 | 5.50 | 13.75 | yes | 13.41 | 0.60 | 4.5 | 5.3 |
| 2 | 3.75 | 12.75 | | | | | |
| 3 | 4.75 | 13.95 | | | | | |
| 4 | 6.75 | 12.85 | | | | | |
| 5 | 6.50 | 12.60 | | | | | |
| 6 | 6.75 | 14.65 | | | | | |
| 7 | 6.75 | 13.35 | | | | | |
| 8 | 5.50 | 14.20 | no | 13.28 | 0.79 | 6.0 | |
| 9 | 6.50 | 13.55 | | | | | |
| 10 | 6.25 | 12.10 | | | | | |

---

[*] Note from the translators: the drop heights given in this table appear inconsistent with those given in the text.

**TABLE 2**

*Formation of bands produced by impact of drop-weights of two different masses dropped from various heights*

| Mass of the drop-weight | Drop height / m | Number of specimens tested | Result | Mass of the drop-weight | Drop height / m | Number of specimens tested | Result |
|---|---|---|---|---|---|---|---|
| 50 kg | 1.5 | 5 | Four specimens had no bands, one had a band inside | 25 kg | 2.55 | 2 | No bands. |
| | 2.0 | 6 | One specimen had no bands, the other five had bands inside | | 3.0 | 2 | One had a band outside, the other had no band. |
| | 2.5 | 11 | One specimen had no bands, the other five had bands outside | | 3.5 | 2 | Both specimens had one outer band |
| | 3.0 | 5 | All five specimens had double bands which exited to the outside | | 4.0 | 2 | Both specimens had one inner band. |
| | 3.5 | 5 | All five specimens had double bands which exited to the outside | | 5.0 | 2 | In addition ot the inner bands, outer bands were also formed |
| | 4.0 | 5 | All five specimens had double bands which exited to the outside | | | | |

*2. Influence of the energy and speed of impact on band formation.*
In the experiments, temperature measurements were performed on ten identical specimens subjected to virtually identical drop-weight impacts. However, bands formed only in seven out of the ten tests. Possible factors on which band formation might depend are the following: impact energy, impact speed, differences in the shape of the specimen, and differences in the axiality of the impacts.

The role of the last two factors seemed unlikely because the dimensions of the specimen were the same to within an uncertainty of 0.2-0.3 mm.

In order to determine the influence of the energy and speed, a series of impacts were performed using different drop heights and two different masses (50 and 25 kg). Moreover the specimens were positioned on the anvils at the imprint of previously tested specimens.

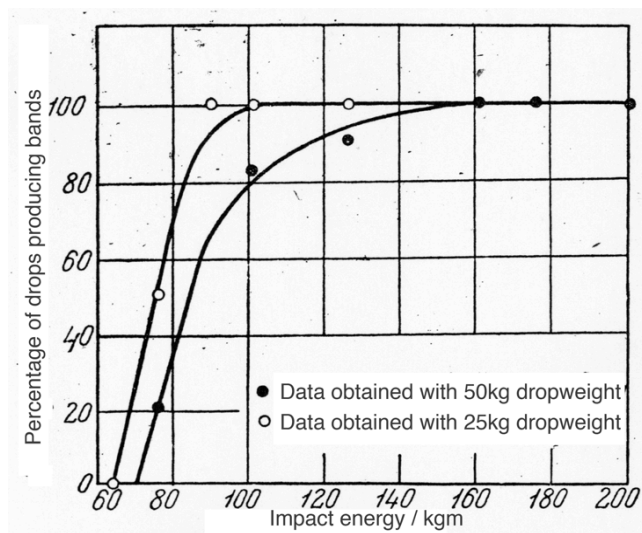

*Fig. 5a. Dependence of the band formation on the impact energy.*

The experiments lead to the following results (see Table 2 and Fig. 5a).

The plots in Fig. 5a were constructed using the information presented in Table 2. The two curves show there is a link between the number of observed bands (expressed as a percentage of the total number of impacts) and the impact energy in kilogram-meters. The two curves are for the two different drop-weights. Despite the small number of tests performed, it can be seen that the curve for the 25 kg weight is above and to the left of the one for the 50 kg weight.

Thus, it seems that band formation under the same conditions most likely depends on impact speed so that for the same impact energy band formation is favoured at higher speeds.

As the impact energy is increased, two bands are typically formed which meet each other at a certain angle as can be seen in the cross-sections shown in Figures 1 & 3.

*3. Relative measurements of the hardness*

In order to clarify to what extent the above hypothesis correctly predicts the metallographic condition of the steel in the band, relative hardness measurements were made. These were performed by means of the Martens scratch sclerometer and the Herbert oscillation pendulum. For comparison, a steel specimen with a carbon content of 0.39% was quenched in the cold water after holding at a temperature of 950-980 °C for 20 minutes. The hardness was compared with the bands, obtained in the same steel by impact. The results obtained with a scratch hardness tester are given in Table 3.

**TABLE 3**
*Martens scratch hardness of the bands*

| Specimen preparation | Scratch widths in 'scales' [*] |
|---|---|
| Output structure of the specimen (sorbite in ferrite) | 281, 274, 274, 280: average 277. |
| Quenched specimen to martensite | 147, 147, 147, 147: average 147. |
| Specimen compressed by a 50 kg drop-weight dropped from a height of 2.55 m. | Outer band: in the lateral direction 126, 127, average 127; in the longitudinal direction 129, 129, 129, 129: average 129 Outside of the band: in the lateral direction: 204, 216, average 210; in the longitudinal direction 236, 200, 220, 240 : average 228. Inside of the band: in the lateral direction 137, 137, 140, average 136. Outside of the band: 228, 230, 227: average 228. |
| Specimen compressed by a 50 kg drop-weight dropped from a height of 3 m. | Inside of the band: in the lateral direction 142, 140, 138, 143,144,146, average 142. Outside of the band: 223, 231,230, 241, 245, 227,230, 233: average 232. |

From the results obtained it can be seen that the hardness of the band exceeded that of martensite for the same steel, but only very slightly; in some cases they were almost the same. It should be noted that the measurements were very difficult to make on the inner band due to their small width; Therefore, larger errors are shown for the hardnesses of the inner band compared to those for the outer band.

It was only possible to use the Herbert pendulum-hardness tester on the outer exit point of the bands. For the purpose of this study, this was sufficient to get the values. The pendulum-hardnesses of the outer exit point of the bands are given in Table 4.

---

[*] 140 scales corresponds to a scratch width of 0.01 mm

**TABLE 4**

*Herbert pendulum hardnesses of the bands*

| Specimen preparation | Period of ten pendulum oscillations / s | |
|---|---|---|
| Specimen quenched to martensite | 69.0, 68.0, 70.8, 70.4, 68.4: Average 69.3 | |
| Specimen compressed by a 50 kg drop-weight dropped from a height of 4 m. | Inside of the band | Longitudinal vibrations: 50, 42.4, 46.2: Average 46.2 Transverse vibrations: 42.2, 43.2, 43.2, 42.6: Average 42.8 |
| | Outside of the band | 39.2, 34.4, 36.0, 32.0, 30.4, 36.6, 34.0: Average 34.4 |
| Control glass specimen | Before the experiment: After the experiment: | 98, 100, 102, 100 100, 100, 100, 98 |

In assessing the data presented in this table, one must note that the outer exit points of the bands were very thin and therefore they could bend under the weight of the pendulum; therefore, the determination of the Herbert hardness could not be measured as accurately.

*4. Determination of the specific gravity of the specimens with and without bands.*
Of great interest was the determination of the density of the new phase of the steel. This could serve as a partial check of our working hypothesis, since an increase of the specific volume is a well-known fact for the martensitic transformation. For this purpose specimens with and without bands were chosen. They had all been subjected to a previous impact, so that the possible influence of the different processing grades on the density could be excluded.

The determination of the density was performed by hydrostatic weighing on an analytical balance. For better accuracy in the results, the specimens were cleaned using carbon tetrachloride before the measurement. Before each specimen was weighed in water, care was taken to remove gas bubbles from them. Also the change in density of water with temperature was taken into account.

To perform the measurement, a specimen with a band was cut into three parts, one of which contained the band.

The specific gravities obtained are shown in Table 5. For comparison in the same table, the change in the specific gravity of the same steel was measured. From the table it is evident that the formation of a band is accompanied by a decrease in the specific gravity.

**TABLE 5**
*Determination of the specific gravity of the specimens.*

| Specimen preparation | Sample parts | Weight in air | Water temperature | Weight in water | Density of water at the given temperature | Specific weight | % decrease in the specific gravity | Mean % decrease of the specific gravity |
|---|---|---|---|---|---|---|---|---|
| Specimen impacted by 50 kg weight from a height of 2.55m | Without band | 0.9870 | 14.0 | 0.8602 | 0.9993 | 7.778 | 0.57 | 0.63 |
|  | With band | 0.8720 | 14.9 | 1.6302[*] | 0.9991 | 7.734 |  |  |
| Specimen impacted by 50 kg weight from a height of 2.55m | Without band | 0.3535 | 17.8 | 0.3091 | 0.9987 | 7.931 | 0.72 |  |
|  | Without band | 0.5439 | 17.8 | 0.4752 | 0.9987 | 7.917 |  |  |
|  | With band | 2.9562 | 18.5 | 2.5893 | 0.9985 | 7.877 |  |  |
| Specimen impacted by 50 kg weight from a height of 3m | Without band | 1.6343 | 18.0 | 1.4272 | 0.9986 | 7.879 | 0.59 |  |
|  | Without band | 1.0809 | 15.5 | 0.9448 | 0.9990 | 7.932 |  |  |
|  | With band | 0.0190 | 18.0 | 5.2541[*] | 0.9986 | 7.858 |  |  |
| Steel specimen with a carbon content of 0.39% that has not been impacted | initial structure | 10.2405 | 16.2 | 8.9340 | 0.9989 | 7.830 | 0.52 |  |
|  | Quenched to martensite | 9.5894 | 16.2 | 8.3596 | 0.9989 | 7.789 |  |  |

---

[*] Note from the translators: these two figures are as given in the original paper, but they cannot be correct. The specimens cannot weigh more when immersed in water.

*5. Production of different types of bands in steels.*
The same upsetting tests were carried out on pure iron (less than 0.05% carbon content), an austenitic steel containing chromium-nickel, and also some carbon-containing steels with carbon contents of 0.39%, 0.66% and 0.96%.

(a) Pure iron: This was subjected to a series of sequential, increasing impacts, but no bands were observed. Micsocropic investigation revealed extraordinary elongation, grinding and blending of the grains. Good visual orientation occurred only in the middle layer of the specimen. Their direction was parallel to the impact plane at places near which the polyhedral shape of the grains remained.

(b) Austenitic steel (0.27-0.35% C, 1.55-1.8% Cr, 2.35-25.5% Ni, 0.85-0.95 Mn): After impact a relatively small amount of austenitic steel formed on the surfaces of the specimens in the form of stripes (swelling), which run crosswise to each other. A further increase in the impact energy causes the tearing to occur (e.g. Fig. 8). The tearing direction corresponds to that of the slip planes which are particularly visible along the edges of the specimen shown in Fig. 8. Microscopic examination of a polished and etched section clearly revealed localized deformation, which has a passing resemblance to the band (Fig. 9).

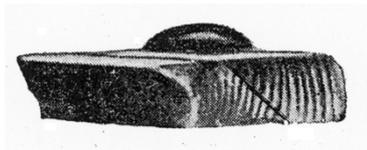

*Fig. 8. A specimen of austenitic steel that was impacted by a 50 kg weight dropped from a height of 2.5 m. On one surface of the specimen, a crack is visible. On the second surface the sliding tracks of the layers can be seen*

It was found that these bands have different properties to the others. First, they etch much more easily than the bulk metal and second the measured Martens hardness was half that of the other bands. The widths of the scratches (in scale divisions) for a load of 50 kg were:

   The initial structure: 225, 232, 232, 228, average 229.
   Inside the band: 215, 207, 210, 216, 221, 206, 216, average 213.
   Outside the band: 225, 227, 226, 238, 236, 237, 239, average 233.

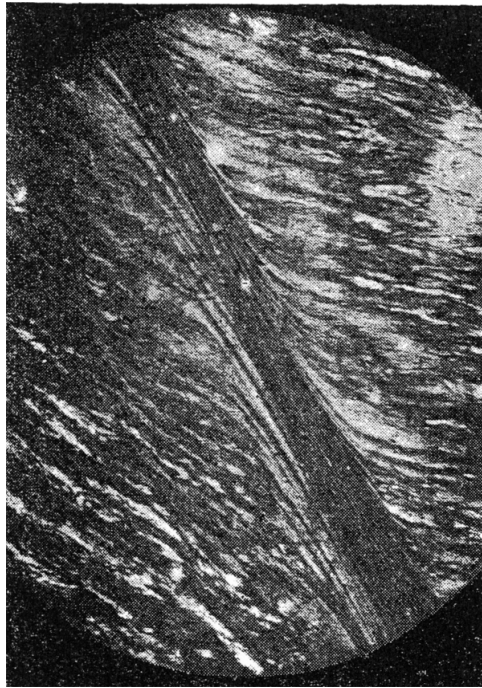

*Fig. 9. Structure of the austenitic steel after impact by a 50 kg weight from a height of 2.5 m.*

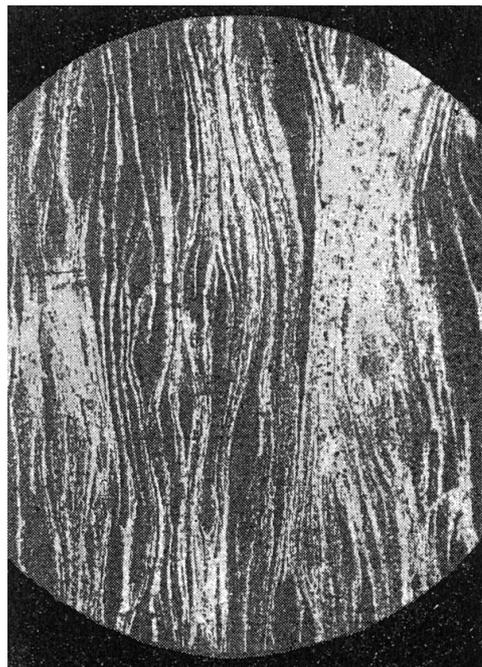

*Fig. 10. Structure after impact of the central part of a specimen with 0.66% C. Initial structure: pearlite and ferrite. Drop-weight mass 50 kg, height 4 m.*

Here, the hardnesses of the initial structure and the band on the outside of the specimen differ little from each other.

(c) Steel with a carbon content of 0.66%. The initial structure of this steel was a regular lamellar pearlite in a ferrite background. This condition was clearly due to imperfect annealing, i.e. annealing at a temperature below the critical point $AC_3$, since pronounced ferrite was missing.

After impact, almost no change in the structure was found in the outer layer of the specimens, while the middle part suffered a heavy deformation (Fig. 10). As for pure iron, the orientation of the elongated grains ran in a direction parallel to the surface of the specimen on which the impact occurred. Even really strong impacts (with a mass of 200 kg) did not cause band formation in this kind of steel.

(d) Steel with a carbon content of 0.96%. Different thicknesses of this steel were impacted, but it did not change its structure: its initial and final state always remained the same, having the form of granular pearlite. No indications of band formation were observed.

*6. Effect of heat treatment on band formation in steel.*
The experiments so far discussed have shown that besides the type of steel, the structural state of the metal also influences band formation. On the one hand, we wanted to determine which of the factors favours easy band formation. On the other hand, it was of interest to obtain a band in steels with a carbon content higher than 0.39%.

To this end, experiments were carried out on steels with carbon contents of 0.39% and 0.66%, which were first annealed to obtain the pearlite in ferrite structure, and then second were processed through quenching and annealing to sorbite.

(a) Annealed steels with 0.39% and 0.66% C were subjected to increasing impacts. In both steels, inner and outer bands were obtained. Microstructural analysis gave very sharp images of slip and branching of the band (Fig. 7). In addition, the bands were relatively densely crosssed by pieces of ferrite. Essentially, there were no differences between the bands in these two steels and those previously obtained .

(b) Steel with 0.39% and 0.66% C, processed to sorbite. It was in these steels that entirely new phenomena were observed. In the steel with 0.66% C, the same kind of cracks, as described previously in austenitic steels, were obtained on the outer plane when a certain impact force was used (see Fig. 9). With increasing impact energy, the specimens split into two along a plane where the band was. Increasing the impact energy still more resulted in the two parts being welded back together.

Sorbite specimens of steel with 0.39% C could be split into parts by any impact, and showed only the usual oriented cracks and bands.

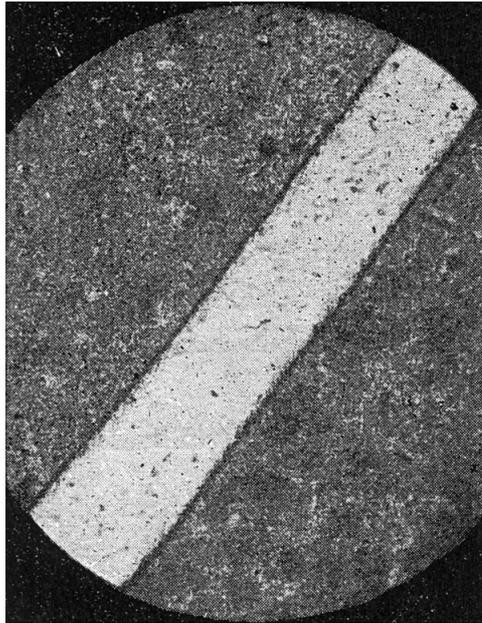

*Fig. 11. A band in a steel specimen with 0.39% C. Initial structure: sorbite. Weight - 50 kg, height - 3.5 m*

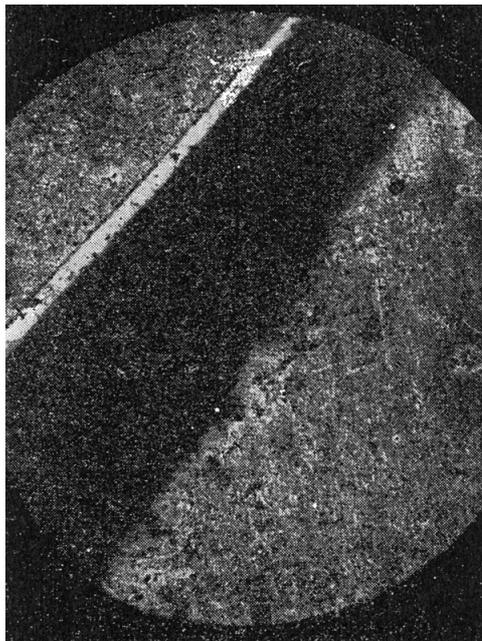

*Fig. 12. A crack, which is a continuation of the band in the same specimen shown in Fig. 11. One edge of the crack was covered with a thin layer of the new phase.*

Microanalysis showed that the plane, along which the 0.66%C steel specimen split was covered with a thin layer of the new phase. A cross-section of the welded specimen showed a sorbite-like basic structure crossing a crack. This was surrounded by the phase substance. From this crack, very thin bands went to many places, these bands being curved in many different ways. In the steel with 0.39% C, a very well-defined band was generated inside the specimen (Fig. 11). Near the edge of the specimen it turned into a crack. The edges of the crack retained a thin layer of the phase (Fig. 12).

*7. Flow of the annealing of the bands.*
From the above, it was not possible to detect the bands' structure, even at very large magnifications (x1200-x2000) but using strong etching with various reagents, it was possible at least to follow microstructural changes during thermal processing.

For this purpose, steel specimens with 0.39% and 0.66% C were heat-treated at 850 °C for half an hour, then annealed and cooled in the furnace. They were then subjected to the same intensity of impact (200 kgm) in which a well-defined band was previously formed.

Under the microscope, the usual shiny band on the base of the pearlite structure in ferrite could be seen (Fig. 7). The specimens were then subjected to annealing at temperatures of 400°, 500°, 600°, 750° and 800°C for 30 minutes, and then cooled in the furnace.

Microstructural examination revealed, that annealing at 400 °C resulted in the troostite-sorbite structure appearing in the band, (Fig. 13). Annealing at 500 °C and 600 °C resulted in the amount of dispersed cementite gradually reducing, and the structure of the band begins to break down. Annealing at 750 °C transformed the structure of the band into granular pearlite (Fig. 14). Microanalysis showed that for specimens annealed at 800 °C, the whole section was perfectly homogeneous, there was nothing distinctive from the main body of the structure which was lamellar pearlite in ferrite (Fig. 15). In both steels with 0.39% and 0.66% C, the change in the structure with increase of the temperature is the same.

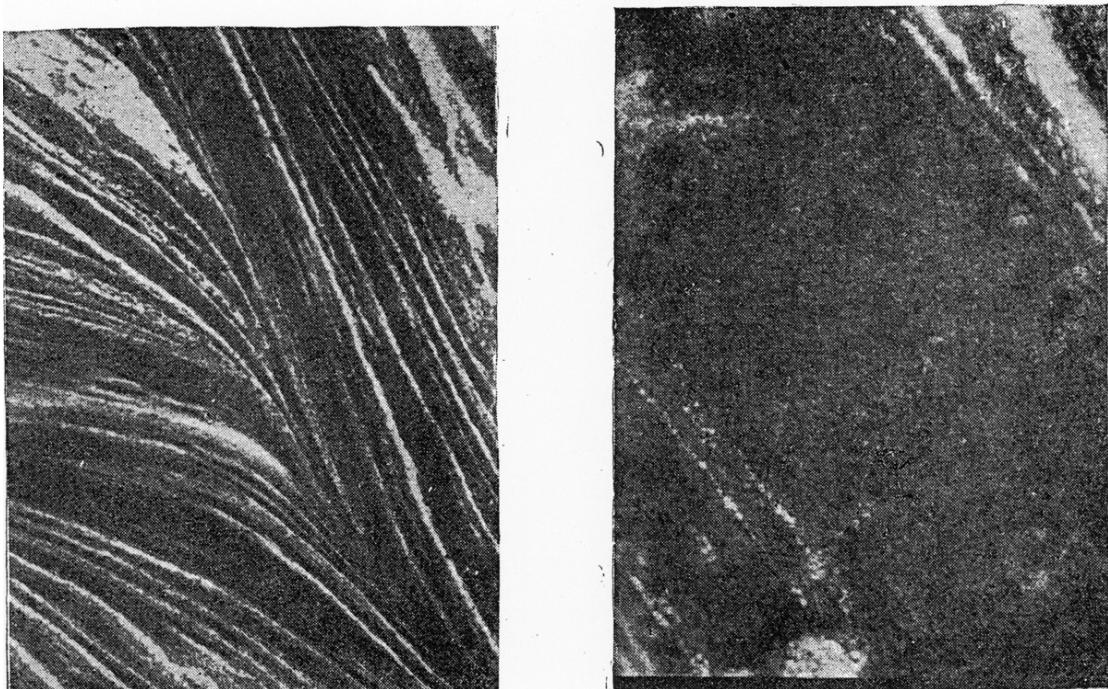

*Fig. 13. Troostite-sorbite structure, one band developed after annealing at 400°C. Steel with 0.66% C; initial microstructure: pearlite in ferrite. Weight – 50 kg, height – 4 m.*

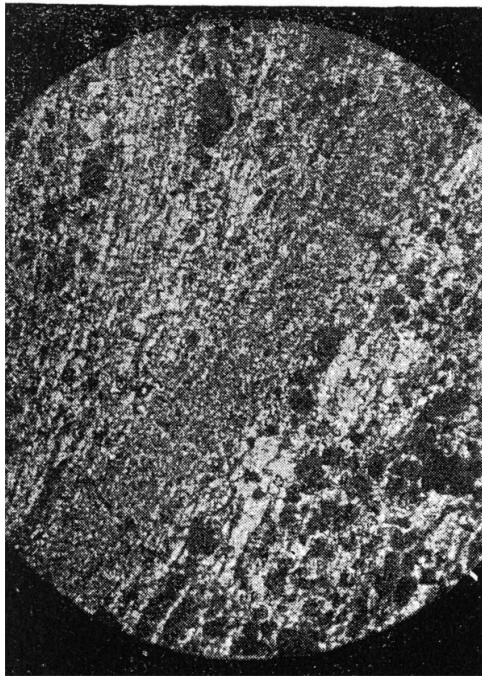
*Fig. 14. Structure of the granular pearlite, one band developed after annealing at 750°C. Steel with 0.39% C; initial microstructure – pearlite and ferrite, weight – 50 kg, height – 4 m.*

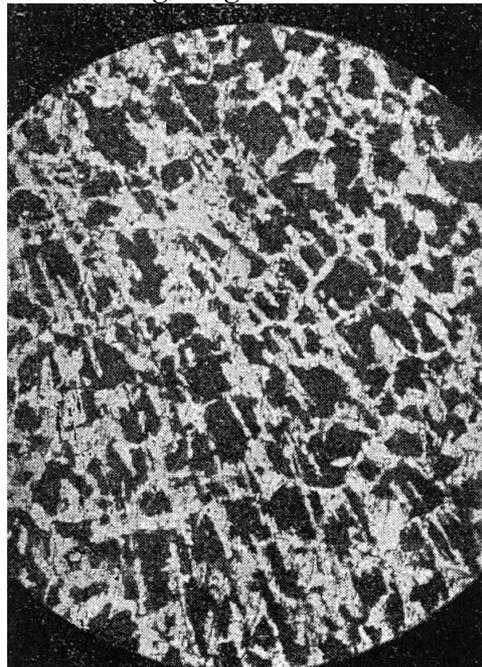
*Fig. 15. Structure of the band after annealing at 800°C .Steel with 0.39% C; initial microstructure – pearlite and ferrite, weight – 50 kg, height – 4 m.*

**Discussion of results**
By comparing the experimental results described above, we reached the following conclusions:

1. Bands form on a plane of localized deformation (slip planes), on which the greatest shear stress develops.
This is supported by: (i) the constant orientation of the band at an angle of between 45° and 55° to the direction of impact; (ii) clearly visible sliding tracks in the microstructure of austenitic steel (Fig. 9); (iii) cracks on the outer surface of austenitic steels (Fig. 8) and in the interior of specimens of sorbitic steel with 0.39% C (Fig. 12) and the splitting of specimens into two for sorbitic steel with 0.66% C.

2. The volume occupied by the band concentrates a significant part of the thermal energy that is generated by mechanical deformation.

3. The scratch hardness of the bands matched to within an order of magnitude the hardness of the martensite produced by quenching a steel of the same composition (Table 3). The slightly greater hardness of the band (142 instead of 147, i.e. 3.5%) must be traced back to the specific conditions of quenching under pressure.

4. Band formation is accompanied by a decrease in the specific gravity of the steel. From Table 5 it is evident that this decrease (0.62%) is very close to 0.52%, which the value for the same steel when quenched to martensite. But here one must take into account that the specific gravity of the metal outside the band does not decrease during impact, but is significantly increased (from 7.83 to 7.89, i.e. by 0.8%), which could only enlarge the differences. This was previously observed in tension by M.G. Oknov (*J. Russ. Metall. Soc.*, no. 1, 1928).

5. The band formation depends not only on the composition of the steel, but also on its condition.

Most clearly, it can be said that the steel that formed the clearest bands was the one having a carbon content of 0.66%. In the sorbite structure, bands were clearly visible. They were less clear in the pearlite in ferrite and were never observed in the structure consisting of individual pieces of pearlite and ferrite. The same can be said about the steel with a carbon content of 0.39%.

The influence of the microstructure also explains why steel specimens with a carbon content of 0.96% never formed bands in these drop-weight experiments.

6. Bands in the austenitic steels differ from those in carbon-containing steels. They were apparently formed by localized deformation in front (ahead?) of the slip planes, which produced strong local heating. However, quenching with transformation to martensite could not occur due to the known properties of austenitic steel. A certain increase in the hardness of the band in comparison with the main body of the specimen could be explained by a purely mechanical effect due to partial transformation of austenite to martensite (as has already been frequently observed), but to a degree that means that transformations that take place during quenching of a carbon steel cannot be compared with it. Also it is not observable under the microscope.

Bands were not observed in pure iron. If they were to be observed, this would negate the hypothesis of the phase transformations. However, this has no decisive experimental importance, since localized deformation was not found in iron specimens.

8. Annealing of the band at different temperatures produces structures which are identical to the structures produced by annealing a quenched steel.

All of these facts lead to the conclusion that the microstructure of the band forms by quenching of steel that has previously undergone very special and unexplored conditions. The same conclusion also comes from X-ray analysis, which is described in the work of N. Selyakov and M. Gen published in the *Journal of the Russian Metallurgical Society*, 1930, no. 9-10.

We wish to remark here that our investigations found no enrichment of carbon in the volume occupied by the band after annealing, nor any local accumulations of ferrite in the vicinity of the band before quenching, in contrast to the observations of Kravz-Tarnavskii.

We put forward two hypotheses to explain the fact that the majority of the deformation is concentrated in a particular layer of the metal and does not extend uniformly over the entire volume.

They are:
1. In the initial stages of impact, shear first occurs in the plane of maximum shear stress. This shear grows under a large normal pressure and is therefore accompanied by considerable internal friction and by large local heating. Heating lowers the yield stress, so hardening is followed by weakening. So once slip begins, further deformation continues along the same plane rather than in other regions by favouring the dissipation of an ever greater amount of heat. A similar phenomenon occurs in the case of necking in a tensile specimen.

2. Under strong impact, the elastic limit or the frictional resistance can be higher than the shear strength. We interpret this as the complete breaking of bonds parallel to the planes of largest shear strain. This is analogous to the effects taking place at the edges of the surface along the constriction of the fractured sample[*]. Then, an inclined crack forms, which eventually will break the specimen into two halves. With continued application of the pressure, the upper part of the specimen starts to slide over the lower part, in the process overcoming a large frictional resistance. The high speed of the process does not allow the evolving heat to diffuse into a large volume, leading to a sharp local increase in the temperature. Thus, the crack welds itself together and the heated metal layer which has undergone austenite transformation is quenched.

Finally, we consider it as a pleasant obligation to express our thanks to V. Kravz-Tarnavskii, M. G. Oknov and N. Y. Selyakov for valuable advice and interest in our work, and V. Kravz-Tarnavskii for participation in the completion of the pilot experiment and provision of the specimens.

Leningrad,
Institute of Physics and Technology
Received on 7 April 1935

---

[*] Note from the translators: the meaning of the German of the previous two sentences (part of one sentence in the original) is not clear. So we have rendered them literally into English.